\documentclass[12pt]{iopart}
	\usepackage{textcomp}
	\usepackage{graphicx} 
	\usepackage{harvard}
\def\be{\begin{equation}}
\def\ee{\end{equation}}

\newcommand{\ket}[1]{|#1\rangle}
\newcommand{\pr}{\partial_\rho}
\newcommand{\pf}{\partial_\varphi}
\newcommand{\stkout}[1]{\ifmmode\text{\sout{\ensuremath{#1}}}\else\sout{#1}\fi}
	\usepackage[breaklinks=true,colorlinks,citecolor=blue,linkcolor=blue,urlcolor=blue]{hyperref}
	\usepackage{bm}
	\usepackage{cite}
	\usepackage{float}
	\usepackage{iopams}
\expandafter\let\csname equation*\endcsname    \relax
\expandafter\let\csname endequation*\endcsname\relax
\usepackage{amsmath}

\begin{document}

\title{Finite-size effects in cylindrical topological insulators}
\author{Michele Governale$^{1}$, Bibek Bhandari$^2$, Fabio Taddei$^3$, Ken-Ichiro Imura$^4$ and Ulrich Z\"ulicke$^1$}
\address{$^1$School of Chemical and Physical Sciences and MacDiarmid
Institute for Advanced Materials and Nanotechnology, Victoria University of
Wellington, PO Box 600, Wellington 6140, New Zealand}
\address{$^2$Scuola Normale Superiore \& NEST, NANO-CNR, Pisa, Italy}
\address{$^3$NEST, NANO-CNR \& Scuola Normale Superiore, Pisa, Italy}
\address{$^4$Department of Quantum Matter, AdSM, Hiroshima University,
Higashi-Hiroshima 739-8530, Japan}
\ead{michele.governale@vuw.ac.nz}

\begin{abstract}
We present a theoretical study of a nanowire made of a three-dimensional topological insulator. The bulk topological insulator is described by a continuum-model Hamiltonian, and the cylindrical-nanowire geometry is modelled by a hard-wall boundary condition. We provide the  secular equation for the eigenergies of the systems (both for bulk and surface states) and the analytical form of the energy eigenfunctions.
We describe how the surface states of the cylinder are modified by finite-size effects. In particular, we provide a $1/R$ expansion for the energy of the surface states up to second order.
The knowledge of the analytical form for the wavefunctions enables the computation of matrix elements of any single-particle operators. In particular, we compute the matrix elements of the optical dipole operator, which describe optical absorption and emission, treating intra- and inter-band transition on the
same footing. Selection rules for optical transitions require conservation of
linear momentum parallel to the nanowire axis, and a change of $0$ or $\pm
1$ in the total-angular-momentum projection parallel to the nanowire axis. The magnitude of the optical-transition matrix elements is strongly affected by the finite radius of the nanowire. 
\end{abstract}
\maketitle

\section{\label{sec:Intro}Introduction}
Three-dimensional (3D) topological insulators (TI)s were predicted in 2007~\cite{fu2007} as electronic systems characterized by an insulating bulk and gapless conducting surface states (for a review, see Refs.~\cite{hasan2010,xiao2011,hasan2011,ando2013}). The states at the interface between the system and the vacuum are topologically protected against time-reversal invariant perturbations and consist, at low energy, of single Dirac fermions \cite{dung09,parente11,imura12}.
Recent advances in nanofabrication techniques have enabled the realization of 3D-TI samples of reduced dimensionality, for example in the form of nanowires\cite{peng2009,xiu2011,hong2012,
tian2013,wang2013,hamdou2013,safdar2013,
hong2014a,Jauregui2015,Cho2015,bassler2015,arango2016,Dufouleur2017,bhattacharyya2017,ziegler2018,Kunakova2018,Kim2019,Munning2019}. 
3D-TI nanowires proximised with an s-wave superconductor have been proposed as a possible platform for the realization of Majorana bound states \cite{Cook2011,deJuan2014}.
The availability of nanometer-scale samples is interesting also because it offers the opportunity to investigate the competition between the inverted bulk gap and the size-quantisation energy as well as the extent of the  localization of surface states\cite{zhou2008,linder2009,liu2010,
zhang2010,bardarson2010,imura2012,
hong2014b,iorio16,kotulla2017,gioia2019}. 
In Ref.~\cite{iorio16}, an approximate analytic model supplemented by a numerical scheme based on exact diagonalisation was introduced to study the quantum interference effects on the low-energy spectrum of $\text{Bi}_2\text{Se}_3$ nanowires.

In this paper we explore the properties of a finite-radius 3D-TI cylinder, using the envelope-function description of the TI bulk band structure developed in Refs.~\cite{zha09,liu10}. Our goal is to determine the dependence of its energy spectrum and eigenfuctions on the radius $R$. 
The central point of our analysis is the analytical expression of the eigenfunctions, which allow us to express cylindrical hard-wall boundary conditions in terms of secular equations that can be approximated in the limit of large radii: we obtain approximate expressions for the eigenenergies up to second order in $1/R$.
The analytical functional form of the eigenfunctions, which is valid irrespective of the radius of the wire, enables the calculation of the matrix elements of any observable. As an example, we consider the dipole matrix elements for optical transitions. In particular, we find that the selection rules for absorption and emission are not modified by a finite radius, in contrast to the case of a spherical nanoparticle\cite{gioia2019}. 
Numerical results are presented for two different materials, namely Bi$_2$Se$_3$ and Bi$_2$Te$_3$, which show qualitatively different behaviours. 
We compute eigenenergies as functions of the radius $R$ and longitudinal momentum and compare them with approximate large-radius expressions. The eigenenergies are found to be oscillating for small values of $R$, especially in the case of Bi$_2$Te$_3$.
Moreover, we characterize the behaviour of eigenfunctions by plotting the average radial coordinate and the corresponding variance as a function of the radius $R$. As expected, the average coordinate moves towards the centre of the nanowire for small values of $R$, more rapidly for Bi$_2$Te$_3$ than for Bi$_2$Se$_3$, while the variance increases in an oscillating fashion for increasing radii, reaching the asymptotic value more rapidly in the case of Bi$_2$Se$_3$ with respect to Bi$_2$Te$_3$.
Finally, we calculate numerically the dependence of the optical dipole matrix elements on the radius finding quantitative important changes with respect to the bulk situation.

The paper is organized as follows. In section~\ref{sec:level1}, we present an analytic treatment for a cylindrical 3D-TI nanowire with hard-wall confinement. We conclude section~\ref{sec:level1} with a complete analytic expression for the eigenfunction of the finite-radius 3D-TI. In section~\ref{sec:results}, we study the finite size effects on the topological properties of a cylindrical 3D-TI for two different materials. Specifically, we study the eigenenergies and characterise the eigenfunctions of the system as a function of the radius of the cylinder.
Finally, in section~\ref{sec:opt}, we calculate the optical dipole matrix elements of a cylindrical TI and study their dependence on the the radius of the cylinder.

\section{Model}
\label{sec:level1}
We consider an infinitely long cylinder of TI of radius $R$, whose axis is in the $z$-direction. The bulk TI is described by the Hamiltonian\cite{zha09,liu10}
\begin{align}
\label{eq:H0}
H_0 &=\left( 
\begin{matrix}
m(\mathbf{p}) & B p_z & 0 & A p_-\\
B p_z & -m(\mathbf{p}) & A p_- & 0\\
0 & A p_+ & m(\mathbf{p}) & -B p_z \\
A p_+ & 0 & - B p_z & -m(\mathbf{p})
\end{matrix}
\right),
\end{align}
where $\mathbf{p}=(p_x,p_y,p_z)$ is the momentum operator, $m(\mathbf{p})=m_0+ m_1 p_z^2+m_2 (p_x^2+p_y^2) $ is the mass term and $p_\pm=p_x\pm i p_y$. The  Hamiltonian (\ref{eq:H0}) is written in the basis $\{\ket{P1_{z}^{+}\uparrow}, \ket{P2_{z}^{-}\uparrow}, \ket{P1_{z}^{+}\downarrow}, \ket{P2_{z}^{-}\downarrow} \}$.
When the sign of $m_0/m_2$ is negative, the material is in the topological insulating phase, causing isolated boundaries to host surfaces states represented by gapless Dirac cones. The coefficients $m_0$, $m_1$ and $m_2$, as well as the coefficients $A$ and $B$ of the linear-momentum terms depend on the material\cite{nec16}. The values of the parameters for the most common TIs are reported in Table \ref{tab:parameters}. 
\begin{table}[h]
\caption{\label{tab:parameters}
Values for parameters in the effective continuum-model Hamiltonian
describing bulk-electronic states of currently available
topological-insulator materials, from Ref \cite{nec16}.} 
\renewcommand{\arraystretch}{1.1}
\centering
\begin{tabular*}{0.62\columnwidth}{l|c|c|r}
\hline \hline \rule{0pt}{2.5ex}
& Bi$_2$Se$_3$  & Bi$_2$Te$_3$  &  Sb$_2$Te$_3$ \\
\hline
$m_0$ (eV) & -0.169& -0.296 & -0.182 \\
$m_1$ (eV\AA$^2$) & 3.353 & 9.258 & 22.136\\
$m_2$ (eV\AA$^2$) & 29.375 &177.355 & 51.320\\
$B$ (eV\AA) & 1.836 & 0.900 & 1.174\\
$A$ (eV\AA) & 2.513 & 4.003 & 3.694\\
\hline \hline
\end{tabular*}
\end{table}
As the system has cylindrical symmetry, it is convenient to express $H_0$ in cylindrical coordinates. Following Imura \emph{et al.} \cite{imura11}, we write the Hamiltonian as a sum of two terms 
\begin{equation}
\label{eq:Hsum}
H_0  =H_\perp + H_\parallel \quad ,
\end{equation}
where 
\begin{subequations}
\begin{align}
\label{eq:Hperp}
H_\perp & = \left(
\begin{matrix}
m_\perp & 0 & 0 & -i A e^{-i \varphi}\pr \\
0 & - m_\perp &  -i A e^{-i \varphi}\pr & 0\\
0 & -i A e^{i \varphi}\pr & m_\perp & 0\\
-i A e^{i \varphi}\pr & 0 & 0 & -m_\perp
\end{matrix}
\right)\\
\label{eq:Hparallel}
H_\parallel & = 
\left(
\begin{matrix}
m_\parallel & B p_z & 0 & - \frac{A}{\rho} e^{-i \varphi}\pf \\
B p_z & - m_\parallel &  -\frac{A}{\rho}  e^{-i \varphi}\pf & 0\\
0 &  \frac{A}{\rho}  e^{i \varphi}\pf  & m_\parallel & -B p_z\\
 \frac{A}{\rho}  e^{i \varphi}\pf & 0 & -B p_z & -m_\parallel
\end{matrix}
\right)
\end{align}
\end{subequations}
and with the mass terms given by the expressions
\begin{align}
m_\perp &= m_0+m_2 \left(-\pr^2 -\frac{1}{\rho} \pr\right)\\
m_\parallel &=- m_2\frac{1}{\rho^2}\pf^2+m_1 p_z^2.
\end{align}
The Hamiltonian $H_0$ commutes both with $p_z$ and with the $z$-component of the total angular momentum $(L_z+\frac{\hbar}{2}\sigma_z) \otimes \tau_0$, where $\tau_0$ is the identity matrix in the orbital
pseudo-spin subspace. In the following, to avoid cluttering the notation, we set $\hbar=1$.
The commutation relations of $H_0$ discussed above suggest the following \textit{Ans\"atz} for the wave function:
\begin{align}
\label{eq:Ansatz1}
\Psi(\rho,\varphi,z)=\frac{e^{i k_z z}}{\sqrt{2\pi}}
\left(\begin{matrix}
\Phi_1(\rho)e^{i(j-\frac{1}{2})\varphi} \\
\Phi_2(\rho)e^{i(j-\frac{1}{2})\varphi}\\
\Phi_3(\rho) e^{i(j+\frac{1}{2})\varphi}\\
\Phi_4(\rho)e^{i(j+\frac{1}{2})\varphi}
\end{matrix}\right),
\end{align}
where $k_z$ is the eigenvalue of $p_z$ and $j$ (half integer) the eigenvalue of the $z$ component of the total angular momentum. Solving the eigensystem requires applying the Hamiltonian Eq.~(\ref{eq:H0}) to the wavefunciton in Eq.~(\ref{eq:Ansatz1}). The calculation is detailed in~\ref{secul}. In order to solve the radial part of the eigensystem, we make further \textit{Ans\"atze} for the $\Phi_i(\rho)$ and rewrite  Eq.~(\ref{eq:Ansatz1}) as 
\begin{align}
\label{eq:Ansatz2}
\Psi(\rho,\varphi,z)=\frac{e^{i k_z z}}{\sqrt{2 \pi}}
\left(\begin{matrix}
c_1 J_{j-\frac{1}{2}}(\kappa\rho) e^{i(j-\frac{1}{2})\varphi} \\
c_2 J_{j-\frac{1}{2}}(\kappa\rho)e^{i(j-\frac{1}{2})\varphi} \\
c_3 J_{j+\frac{1}{2}}(\kappa\rho)e^{i(j+\frac{1}{2})\varphi}\\
c_4 J_{j+\frac{1}{2}}(\kappa\rho)e^{i(j+\frac{1}{2})\varphi}
\end{matrix}\right),
\end{align}
where $J_n(z)$ is a Bessel function of the first kind and $\kappa$ and the coefficients $c_1,\dots, c_4$ need to be determined. 
In order for the \textit{Ans\"atz} of Eq.~(\ref{eq:Ansatz2}) to be an eigenfunction of $H_0$ with energy $E$, the parameter $\kappa$ needs to take one of the following two values
\begin{equation}
\kappa_\pm=\Bigg[-\left(\frac{m_0}{m_2}+\frac{A^2}{2 m_2^2}+\frac{m_1}{m_2} k_z^2\right)
\pm
\sqrt{\frac{A^4}{4 m_2^4}+\frac{E^2}{m_2^2}+\frac{A^2 m_0}{m_2^3}+\left(\frac{A^2}{m_2^2}\frac{m_1}{m_2} -\frac{B^2}{m_2^2}\right)k_z^2}\,\Bigg]^{1/2}.
\end{equation}
For the coefficients $(c_1,c_2,c_3,c_4)^{T}$ there are four independent solutions (two for $\kappa_+$ and two for $\kappa_{-}$) given by
\begin{align}
\left(
\frac{i A \kappa_\pm}{\Delta_\pm},\
0,\ \frac{ B k_z}{\Delta_\pm},\ 1
\right)^{T},\,\,
\left(
-\frac{B k_z}{\Delta_\pm},\ 
1,\ -\frac{i A \kappa_\pm}{\Delta_\pm},\ 0
\right)^{T},
\end{align}
where
 $\Delta_\pm=m_2\kappa_\pm^2+m_1 k_z^2+m_0-E$. 
 The general solution for the wavefunction with  quantum numbers $k_z$, $j$ and $E$ is a linear combination of the four independent solutions obtained above:
\begin{equation}
\label{eq:wavefunction}
\Psi(\rho,\varphi,z)=\frac{e^{i k_z z}}{\sqrt{2\pi}}\sum_{s=\pm}\left\{
\alpha_s\left(
\begin{matrix}
\frac{i A \kappa_s}{\Delta_s}\ J_{j-\frac{1}{2}}(\kappa_s\rho)\ e^{i(j-\frac{1}{2})\varphi}\\
0\\ \frac{ B k_z}{\Delta_s}\ J_{j+\frac{1}{2}}(\kappa_s\rho) \ e^{i(j+\frac{1}{2})\varphi}\\ J_{j+\frac{1}{2}}(\kappa_s\rho) \ e^{i(j+\frac{1}{2})\varphi}
\end{matrix}\right)+\beta_s \left(
\begin{matrix}
- \frac{ B k_z}{\Delta_s}\ J_{j-\frac{1}{2}}(\kappa_s\rho)\ e^{i(j-\frac{1}{2})\varphi}\\
J_{j-\frac{1}{2}}(\kappa_s\rho)\ e^{i(j-\frac{1}{2})\varphi}\\ -\frac{ i A \kappa_s}{\Delta_s}\ J_{j+\frac{1}{2}}(\kappa_s\rho) \ e^{i(j+\frac{1}{2})\varphi}\\0
\end{matrix}\right)
\right\}.
\end{equation}
We can now solve the confinement problem by assuming a hard-wall cylindrical confinement potential of radius $R$. We need to impose the boundary condition $\Psi(R,\varphi,z)=0$. This leads to as system of equations for the coefficients $\alpha_s$ and $\beta_s$ which has non-trivial solutions for energies obeying the secular equation
\begin{equation}
\frac{T_j(\kappa_{+}R)}{T_j(\kappa_{-}R)}+ \frac{T_j(\kappa_{-}
R)}{T_j(\kappa_{+}R)} = 
 \frac{\kappa_+\, \Delta_{-}}{\kappa_-\, \Delta_{+}} + \frac{\kappa_-
\, \Delta_{+}}{\kappa_+\, \Delta_{-}} + \frac{B^2}{A^2} k_z^2\, \frac{(\Delta_{+} -
\Delta_-)^2}{\kappa_+ \kappa_- \Delta_+ \Delta_-}\quad,
\label{seceq2}
\end{equation}
where we have defined the function $T_j(z)=\frac{J_{j+1/2}(z)}{J_{j-1/2}(z)}$. 
A detailed derivation of the secular equation is provided in~\ref{secul}.
In the case $k_z=0$, the problem decouples in two $2\times 2$ problems and we have two independent secular equations
\begin{subequations}\label{exactkz0}
\begin{align}
\label{eq:k0sec1}
\frac{\kappa_{+}\Delta_{-}}{\kappa_{-}\Delta_{+}}=
\frac{T_j(\kappa_{+}R)}{T_j(\kappa_{-}R)} \quad ,\\
\label{eq:k0sec2}
\frac{\kappa_{+}\Delta_{-}}{\kappa_{-}\Delta_{+}} =
\frac{T_j(\kappa_{-}R)}{T_j(\kappa_{+}R)} \quad ,
\end{align}
\end{subequations}
which are analogous to Eq.~(28) of Ref.~\cite{gioia2019}. 
The $k_z=0$ energy eigenstates associated with solutions of
Eq.~(\ref{eq:k0sec1}) have $\beta_s=0$ and therefore their only nonvanishing spinor components are the first and the fourth. Conversely, the eigenstates
corresponding to solutions of Eq.~(\ref{eq:k0sec2}) have $\alpha_s=0$ and
therefore their only nonvanishing spinor components are the second and
the third. Taking into account the transformation properties of the basis
states under spatial inversion, it is straightforward to show that
eigenstates associated with energy eigenvalues arising from the secular
equation (\ref{eq:k0sec1}) [(\ref{eq:k0sec2})] are also parity eigenstates
with eigenvalue $(-1)^{j-\frac{1}{2}}$ [$(-1)^{j+\frac{1}{2}}$]. 

Even for
finite $k_z$, the spinors multiplied by $\alpha_s$ [$\beta_s$] in the
\textit{Ans\"atz} (\ref{eq:wavefunction}) remain parity eigenstates with
eigenvalue $(-1)^{j-\frac{1}{2}}$ [$(-1)^{j+\frac{1}{2}}$]. However, as the
energy eigenstates for nonzero $k_z$ are superpositions of these
opposite-parity spinors, they are not eigenstates of parity.

Once we fix the quantum number $j$ and $k_z$ and solve the secular equation (\ref{seceq2}) we obtain a series of solutions both with positive and negative energies. Of these, we will only consider the two, one positive and one negative, with the smallest absolute value of the energy.
 We will indicate the positive(negative)-energy solution with $s=+(-)$. \footnote{In principle, we could introduce another integer quantum number to label the different solutions as in the case of a particle in a box.}
Furthermore, we will restrict our analysis to energies that lie within the bulk gap. The quantum numbers that we will use to label the states are $s=\pm,j,k_z$.
The secular problem yields the full knowledge of the eigenfunctions. In order to simplify the notation, in the following we rewrite the eigenfunction 
Eq.~(\ref{eq:wavefunction}) as 
\begin{align}
\Psi_{s,j,k_z}(\rho,\varphi,z)=\frac{e^{i k_z z}}{2\pi}
\left(\begin{matrix}
\Phi_{1,s,j,k_z}(\rho)e^{i(j-\frac{1}{2})\varphi} \\
\Phi_{2,s, j,k_z}(\rho)e^{i(j-\frac{1}{2})\varphi}\\
\Phi_{3,s, j,k_z}(\rho) e^{i(j+\frac{1}{2})\varphi}\\
\Phi_{4,s, j,k_z}(\rho)e^{i(j+\frac{1}{2})\varphi}
\end{matrix}\right),
\end{align}
where the wavefunction obeys the normalisation condition $\sum_{i=1}^4 \int_0^R  d\rho \rho \left|\Phi_{i,s,j,k_z}(\rho)\right|^2=1$.

\section{Results}
\label{sec:results}
In order to understand the effect of a finite radius of the cylinder and how it affects the topologically protected surface states, we start from the large-radius limit.

\subsection{Large-radius expansion}
A natural length scale in this context is the effective Compton length $R_{0}=\left|\frac{A}{m_0}\right|$. In the following we perform an expansion in $R_0/R$ and find corrections to the asymptotic (large $R$) results obtained by  Imura \textit{et al.}\cite{imura11}.  
To this aim, we make use of Hankel's asymptotic expansion for the Bessel function\cite{abr64}
\begin{equation}
\label{eq:hankel}
J_n(z)\approx\sqrt{\frac{2}{\pi z}}\bigg[ P(n,z)\cos\left(z-\frac{1}{2}n\pi-\frac{1}{4}\pi\right)
-Q(n,z) \sin\left(z-\frac{1}{2}n\pi-\frac{1}{4}\pi\right)\bigg].
\end{equation} 
The functions $P(n,z)$ and $Q(n,z)$ are power serieses of $1/z$. 

\subsubsection{Zero axial momentum}
We start by considering the case of zero axial momentum ($k_z=0$), with the goal to understand the $j$-dependence of the surface states.  We will consider only one of the two secular equations, namely Eq.~(\ref{eq:k0sec1}) which can be recast as
\begin{equation}
 \kappa_+ \Delta_- J_{j-\frac{1}{2}}(\kappa_+ R)J_{j+\frac{1}{2}}(\kappa_- R)
 - \kappa_- \Delta_+ J_{j+\frac{1}{2}}(\kappa_+ R)J_{j-\frac{1}{2}}(\kappa_- R)=0.
\end{equation}
For realistic materials, see Table \ref{tab:parameters}, and small values of energies $E\ll |m_0|$, $\kappa_\pm=k\pm i q$ with $q>0$. 
In the large-radius limit $q R \gg 1$, we keep only the terms proportional to $\exp(q R)$ in Eq.~(\ref{eq:hankel}). 
The secular equation reduces to 
\begin{align}
 &\kappa_{+}\Delta_{-}\left[P\left(j-\frac{1}{2},\kappa_{+}R\right)-iQ\left(j-\frac{1}{2},\kappa_{+}R\right)\right]\left[P\left(j+\frac{1}{2},\kappa_{-}R\right) + iQ\left(j+\frac{1}{2},\kappa_{-}R\right)\right]
 \nonumber
 = \\
 \label{eq:expansion}
&{-}\kappa_{-}\Delta_{+}\left[P\left(j-\frac{1}{2},\kappa_{-}R\right){+}iQ\left(j-\frac{1}{2},\kappa_{-}R\right)\right]\left[P\left(j+\frac{1}{2},\kappa_{+}R\right)-iQ\left(j+\frac{1}{2},\kappa_{+}R\right)\right].
\end{align}
Taking the zeroth order of the Hankel's expansion [i.~e. $P(n,z)=1$ and $Q(n,z)=0$], the secular equation becomes
\begin{equation}
\kappa_+ \Delta_-  {+} \kappa_- \Delta_+=0. 
\end{equation}
This equation has a zero-energy solution if $
m_0/m_2<0$, i.~e. when the system is in the topological phase. 

Next, we consider the next two terms in the Hankel's expansion, that is $P(n,z)=1-{(4n^2-1)(4n^2-9)}/{(128z^2)}$ and $Q(n,z)=(4 n^2-1)/(8z)$, and insert them into Eq.~(\ref{eq:expansion}). 
After some tedious but otherwise standard algebra, we obtain the eigenenergies up to second-order in $R_0/R$ 
\begin{align}
E=A \frac{j}{R}-\frac{A^2}{2 m_0}\frac{j}{R^2}.
\label{1overr}
\end{align}
The first term is in agreement with Ref.~\cite{imura11}, the second term gives the first correction to the asymptotic result. The other solution, with the opposite sign, $E=-A j/R +\frac{A^2}{2 m_0}\frac{j}{R^2}$ arises from solving Eq.~(\ref{eq:k0sec2}).

\subsubsection{Finite axial momentum}
In this section we assume that $k_z R\gg 1$.
Proceeding in the same way as for  case $k_z=0$, in zeroth-order in $R_0/R$ the secular equation for the case of non-zero axial momentum reduces to
\begin{align}
\left(\kappa_+\Delta_{-} + \kappa_-\Delta_+\right)^2+ \frac{B^2}{A^2} k_z^2\left(\Delta_{+}-\Delta_{-}\right)^2=0.
\end{align}
This equation has the solutions 
\begin{align}
E=\pm B k_z,
\end{align}
which represents the linear dispersion of the surface modes. 

Considering the  Hankel's expansion up to terms in $1/z^2$, that is $P(n,z)=1-{(4n^2-1)(4n^2-9)}/{(128z^2)}$ and $Q(n,z)=(4 n^2-1)/(8z)$, we obtain 
the eigenenergies up to second order in $R_0/R$ 
\begin{align}
E=\pm \left(B k_z+\frac{1}{2} \frac{A^2 j^2}{B k_z R^2}\right),
\label{second}
\end{align}
which corresponds to the Taylor expansion in second order in $1/(k_z R)$ of the result by Imura \textit{et al.}\cite{imura11}, $E=\pm \sqrt{B^2 k_z^2+A^2 j^2/R^2}$. Notice that we are not allowed to take the $k_z\rightarrow 0$ limit, as this result has been derived assuming $k_z\gg 1/R$.

\subsection{Numerical results}
\label{numerical}
\begin{figure}
	\centering
	\includegraphics[width=\columnwidth]{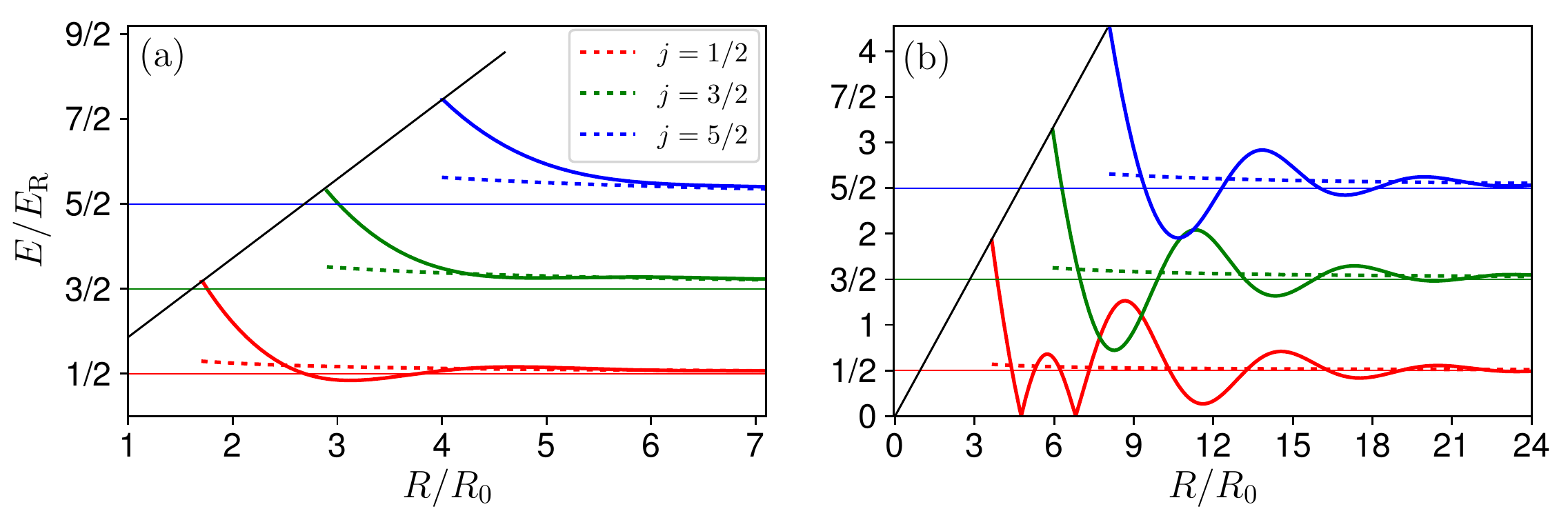}
\caption{Eigenenergies in units of $E_R=A/R$ for a cylinder of (a) ${\rm Bi}_2{\rm Se}_3$ and (b) ${\rm Bi}_2 {\rm Te}_3$ as a function of the radius $R$ for $k_z=0$. We only show the positive energies, i.~e. $s=+$. Solid curves represent the numerical solution of Eqs.~(\ref{exactkz0}), whereas dashed curves represent the approximate large-radius result given in Eq.~(\ref{1overr}).  Due to finite-size effects, for small and decreasing values of $R$ the eigenenergies  increase. For the model under consideration the bulk gap is given by $\text{min}\left[|m_0|,\, \sqrt{-\frac{A^2}{m_2}\left(m_0+\frac{1}{4}\frac{A^2}{m_2}\right)}\right]$, if the square root is real and by $|m_0|$ otherwise. Both for ${\rm Bi}_2{\rm Se}_3$ and ${\rm Bi}_2 {\rm Te}_3$  the bulk gap is given by $ \sqrt{-\frac{A^2}{m_2}\left(m_0+\frac{1}{4}\frac{A^2}{m_2}\right)}$ and is indicated by a black solid line.}
	\label{fig:evsr}
\end{figure}
In this section we present numerical results for two different materials, namely Bi$_2$Se$_3$ and Bi$_2$Te$_3$, using the parameters of Table~\ref{tab:parameters}.
We use the following units for length and momentum,  respectively,
\begin{align*}
R_{0}=\left|\frac{A}{m_0}\right|\quad
\text{and}\quad k_0=\left|\frac{m_0}{B}\right|,
\end{align*}
where $R_0=1.49\,{\rm nm}$ for Bi$_2$Se$_3$ and $1.35\,{\rm nm}$ for Bi$_2$Te$_3$.

\begin{figure}
	\centering
	\includegraphics[width=\columnwidth]{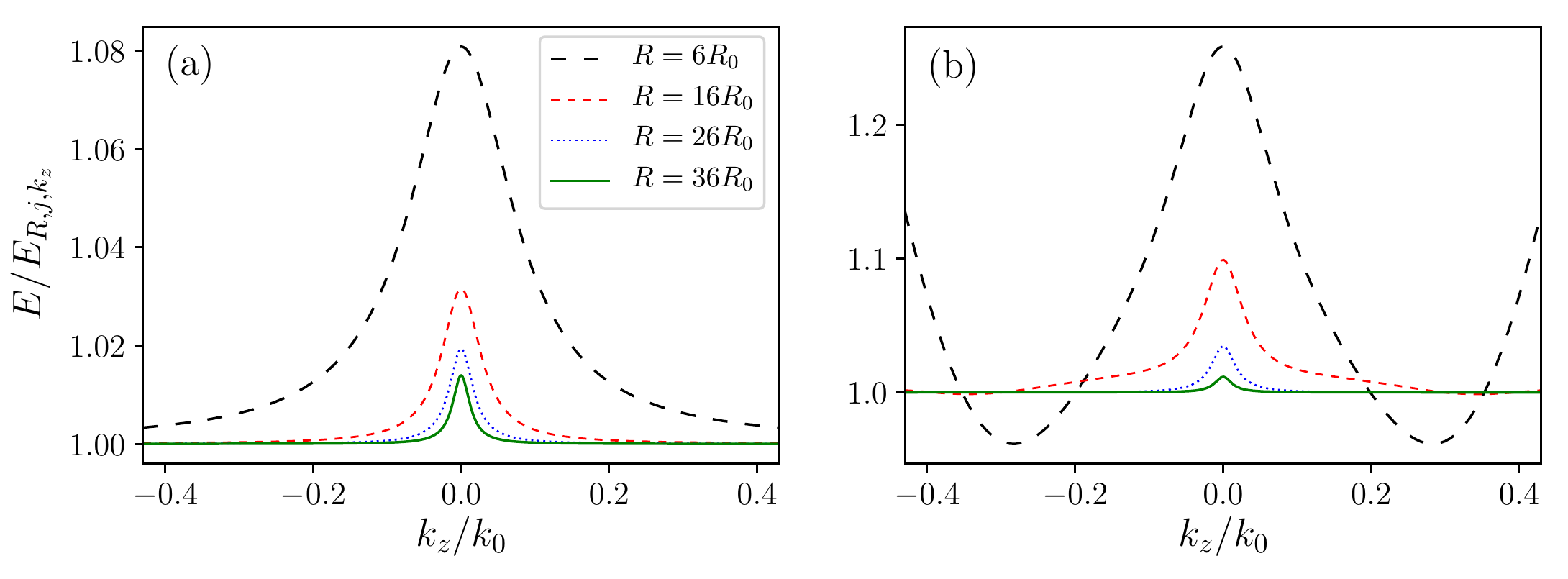}
\caption{Eigenergies divided by the asymptotic value $E_{R,j,k_z}=\sqrt{B^2 k_z^2+A^2 j^2/R^2}$ as a function of wavevector $k_z$ for a cylinder of (a) $\text{Bi}_2\text{Se}_3$ and (b) $\text{Bi}_2 \text{Te}_3$, for $j=\frac{1}{2}$ and different values of radius. In the case of $\text{Bi}_2\text{Te}_3$, for $R=6R_0$ the maximum value of $k_z$ that yields a solution for surface states corresponds to $k_z=\pm 0.43k_0$ where $E=1.14 E_{R,j,k_z}$.}
	\label{fig:evskz}
\end{figure}

Figure~\ref{fig:evsr} shows how the eigenenergies in units of $E_{R}=A/R$ depend on the radius of the cylinder for the two materials and for three different values of $j$. Here we show only the positive energies, that is $s=+$. Solid curves refer to the exact result obtained by solving Eqs.~(\ref{exactkz0}), while the dashed curves  refer to the large-radius analytic expression Eq.~(\ref{1overr}). 
We observe that the latter solutions approximate well the numerical results when $R\gtrsim 6R_0$ for  Bi$_2$Se$_3$, and when $R\gtrsim 20R_0$ for Bi$_2$Te$_3$, respectively.
For Bi$_2$Se$_3$ it is worthwhile noticing that at $R=6R_0$, especially for $j=3/2$ and 5/2, the normalized eigenenergies have not yet reached the asymptotic ($R\gg R_0$) value [represented by the thin solid lines, see Eq.~(\ref{1overr})].
On the other hand, when the radius of the cylinder is small, Fig.~\ref{fig:evsr} shows an oscillatory behaviour, especially in the case of Bi$_2$Te$_3$, that is more pronounced for smaller values of $j$, similarly to a spherical nanoparticle~\cite{gioia2019}. For Bi$_2$Te$_3$, the effect of these oscillations are so large that, for some values of the radius, the surface-state energy goes to zero. This oscillatory behaviour is a consequence of the fact that the wavefunction is no longer localized on the surface of the cylinder. The oscillations are consistent with the results of Ref.~\cite{iorio16} (see also \ref{small}).

In Fig.~\ref{fig:evskz} we show the positive eigeneregies, divided by the asymptotic value $E_{R,j,k_z}=\sqrt{B^2 k_z^2+A^2 j^2/R^2}$, as a function of wavevector $k_z$. Finite-size effects appear in this plot as deviations from unity of the normalized eigenenergies and are more pronounced form small values of $k_z$.

Since we have the full knowledge of the eigenfunctions, we can calculate the expectation values of any single-particle operator. 
The average of the radial coordinate in the state $\Psi_{s,j,k_z}$ is simply given by
\begin{align}
\label{eq:avgrho}
\langle\rho\rangle_{s, j,k_z}=\sum_{i=1}^4\int_0^R d\rho\,2\pi \rho^2 \left|\Phi_{i,s,j,k_z}(\rho)\right|^2,
\end{align}
and its variance by
\begin{equation}
\mathcal{D}\rho_{s,j,k_z}=\sqrt{\left\langle \rho^2\right\rangle_{s,j,k_z}-\left\langle \rho \right\rangle_{s,j,k_z}^2} .
\end{equation}

\begin{figure}
	\centering
	\includegraphics[width=\columnwidth]{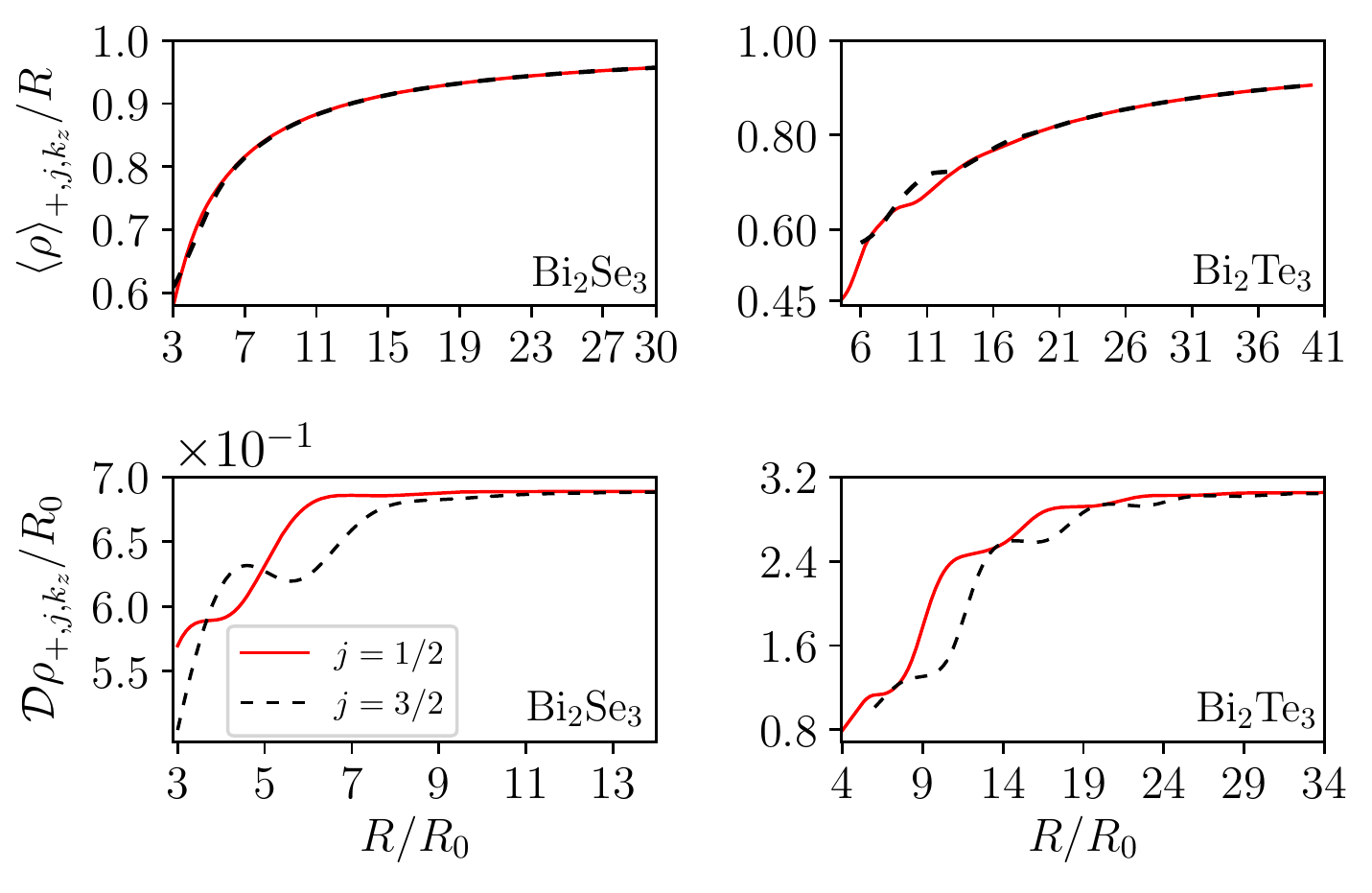}
\caption{Average of the radial coordinate and the corresponding variance versus the radius of a cylinder of $\text{Bi}_2\text{Se}_3$ (left panels) and $\text{Bi}_2 \text{Te}_3$ (right panels) for $k_z=0$ and $j=\frac{1}{2}$ (solid line) and $j=\frac{3}{2}$ (dashed line).
}
	\label{fig:avgrhovar}
\end{figure}

Figure~\ref{fig:avgrhovar} (top panels) shows that the average of the radial coordinate, $\left\langle\rho\right\rangle_{s,j,k_z}$,  approaches $R$ for large values of the radius as expected for topologically-protected surface states.
The average of the radial position for both materials increases monotonically with the radius of the cylinder, showing weak oscillations only for the case of Bi$_2$Te$_3$.
As shown in Fig.~\ref{fig:avgrhovar} (bottom panels), the variance in itself approaches, in an oscillatory fashion, a constant value of the order of $R_0$ for large values of radius (the variance varies very little for $R\gtrsim 8R_0$ for $\text{Bi}_2\text{Se}_3$ and $R\gtrsim 24R_0$ for $\text{Bi}_2\text{Te}_3$).
Since the value of $R_0$ is similar for the two materials ($R_0= 1.5$ nm for Bi$_2$Se$_3$ and $R_0= 1.35$ nm for Bi$_2$Te$_3$), we can conclude that in $\text{Bi}_2\text{Se}_3$ the asymptotic form of the surface states is reached for smaller values of the radius compared to $\text{Bi}_2\text{Te}_3$. 

\subsection{Optical transitions in cylindrical topological insulators}
\label{sec:opt}
\begin{figure}
	\includegraphics[width=\columnwidth]{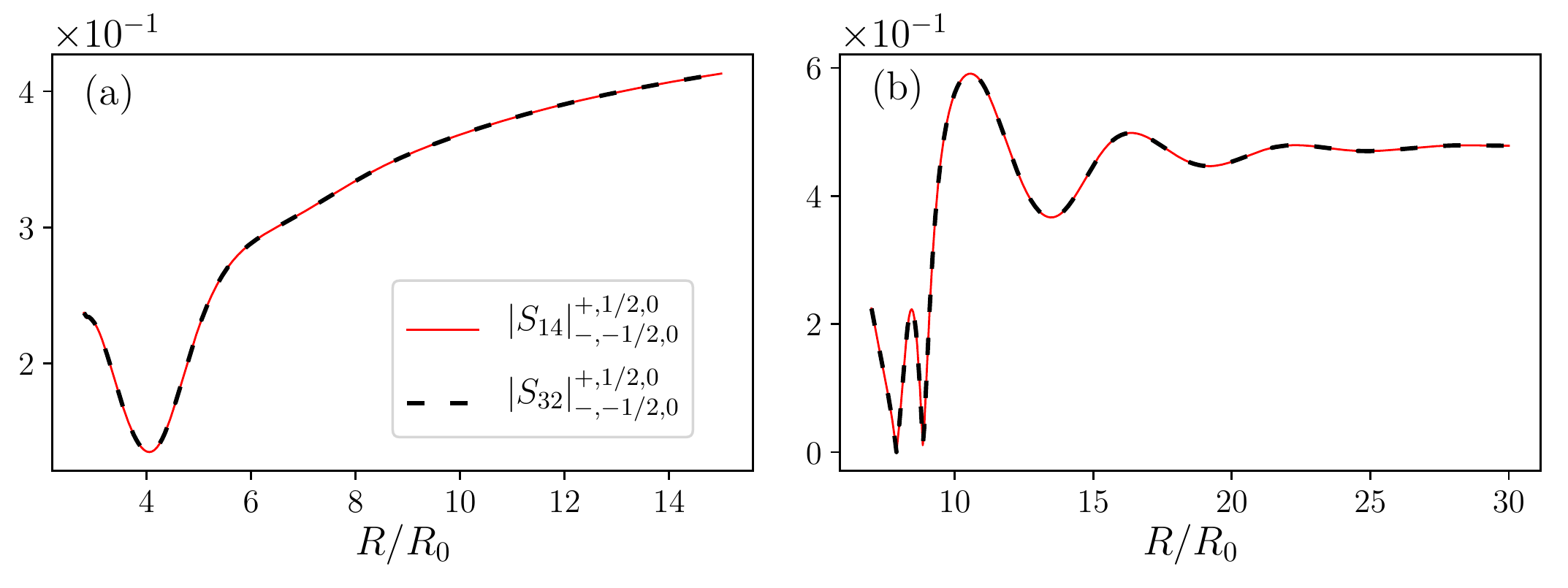}
\caption{Overlap integrals entering the dipole matrix element relevant for absorption for circularly-polarised light as a function of the radius of the cylinder of (a) $\text{Bi}_2\text{Se}_3$ and (b) $\text{Bi}_2\text{Te}_3$ for $k_z=k_z'=0$. We observe that, $|S_{41}|_{-,1/2,0}^{+,-1/2,0}=|S_{23}|_{-,-1/2,0}^{+,1/2,0}=0$ for all values of $R$ considered.  For panel (a),
the smallest value of radius considered is the one corresponding to $E\simeq |m_0|$. For panel (b), instead, the smallest radius considered is the one for which the eigenenergy, given by the solution of Eq.~(\ref{eq:k0sec1}), passes through zero (corresponding to the kink at $R\simeq 7R_0$ in Fig.~\ref{fig:evsr}). }
	\label{fig:dipole1}
\end{figure}

\begin{figure}[!htb]
	\centering
	\includegraphics[width=\columnwidth]{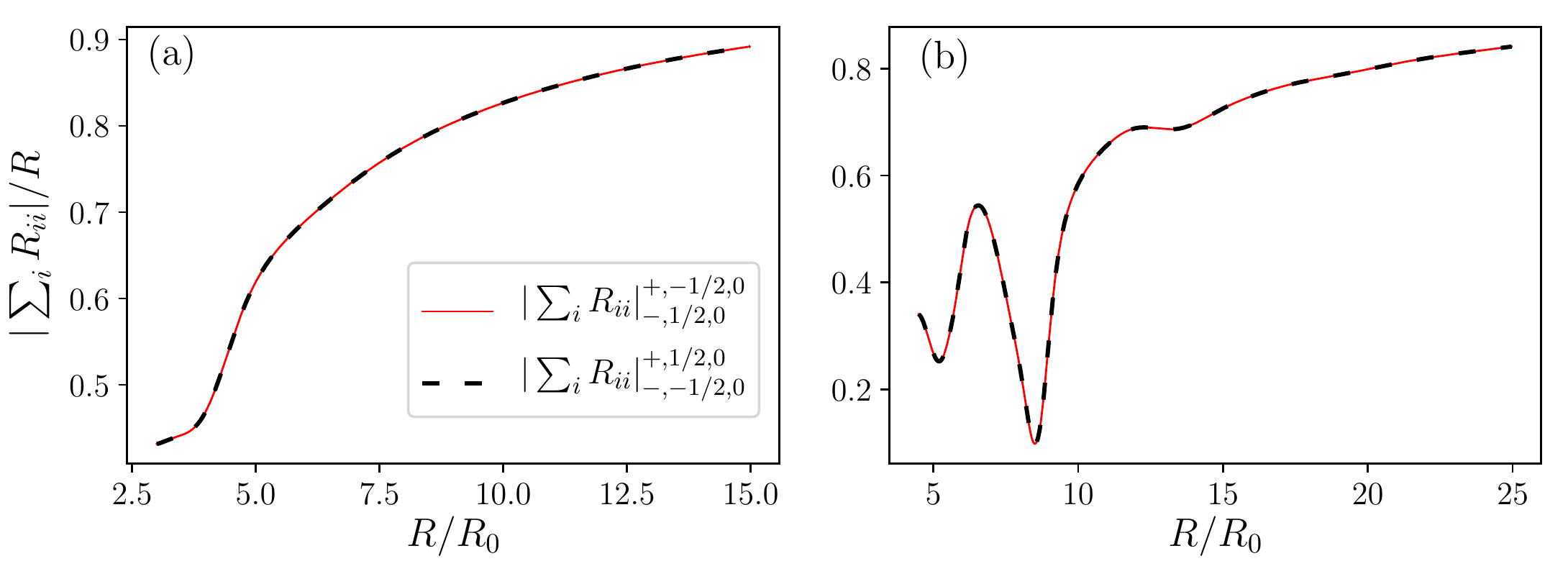}
\caption{Dependence of $\sum_{i=1}^4\mathcal {R}_{ii}$ on the radius of the cylinder of  (a) $\text{Bi}_2\text{Se}_3$ and (b) $\text{Bi}_2\text{Te}_3$ for $k_z=k_z'=0$.}
	\label{fig:dipole2}
\end{figure}

The Hamiltonian in Eq.~(\ref{eq:H0}) can be written as:
\begin{equation}
H_0=\left(m_0+m_1k_z^2+m_2(k_x^2+k_y^2)\right)\tau_z \otimes \sigma_0 
+ Bk_z\tau_x\otimes \sigma_z+A k_x \tau_x \otimes \sigma_x + A k_y \tau_x\otimes \sigma_y,
\label{hamilsim}
\end{equation}
where $\sigma_i$ and $\tau_i$ are Pauli matrices in spin and
orbital-pseudo-spin space, respectively.  
The velocity operator $\mathbf{v}=\partial H_0/\partial{\mathbf{k}}|_{\mathbf{k}=0}$ for the  the Hamiltonian (\ref{hamilsim}) and the optical dipole operator $\mathbf{d}=e\mathbf{r}$ are connected by the fundamental relation\cite{haug2009}:
\begin{equation}
\mathbf{d}_{\tau\sigma,\tau'\sigma'}=\tau\frac{ie}{2m_0}\mathbf{v}_{\tau\sigma, \tau^\prime\sigma^\prime},
\end{equation}
where $\mathbf{d}_{\tau\sigma,\tau'\sigma'}=\left\langle\tau'\sigma'|e\mathbf{r}|\tau\sigma\right\rangle$ and $\mathbf{v}_{\tau\sigma,\tau'\sigma'}=\left\langle\tau'\sigma'|\mathbf{v}|\tau\sigma\right\rangle$.
Here $|\tau\sigma\rangle$ represents the basis functions in the orbital and spin  space of the Hamiltonian $H_0$ defined in Eq.~(\ref{eq:H0}), and $\tau$ is  the eigenvalue of $\tau_z$ associated with the eigenstate $|\tau \sigma\rangle$.
Following the procedure of Ref.~\cite{gioia2019}, we straightforwardly obtain
\begin{equation}\label{eq:dipole}
\mathbf{d}= e\mathbf{r}\,\tau_0\otimes \sigma_0 + \frac{eB}{2m_0}\, \tau_y\otimes \sigma_z\,\hat{z}+\frac{eA}{2m_0}\,  \tau_y\otimes \sigma_y\,\hat{y}+ \frac{eA}{2m_0}\, \tau_y\otimes \sigma_x\, \hat{x},
\end{equation}
for the optical-dipole operator in envelope function representation. The
expression given in Eq.~(\ref{eq:dipole}) accounts on the same footing for both
envelope-function-mediated (sometimes called \emph{intraband}\/) transitions,
which are associated with the first term on the r.h.s., and
basis-function-mediated (\emph{interband}\/) transitions, which are subsumed
in the remaining three terms. The optical dipole matrix elements are given by
\begin{equation}
\nonumber
\mathbf{d}^{s',j^{\prime}k_z'}_{s,j,k_z}=
\int dz\int_0^R  \rho\; d\rho \int_0^{2\pi} d\varphi\
\Psi_{s',j',k_z'}^\dagger(\rho,\varphi,z)\, \mathbf{d}(\rho,\varphi,z) \,\Psi_{s,j,k_z}(\rho,\varphi,z).
\end{equation}
Using Eq.~(\ref{eq:wavefunction}) and performing the integrals over $\varphi$ and $z$, we obtain
\begin{equation}
(d_x+i d_y)^{s',j^\prime,k_z'}_{s,j,k_z}=\delta_{k_z,k_z^\prime}\delta_{j',j+1}\Big[e \sum_{i=1}^{4}\left(\mathcal{R}_{ii}\right)^{s',j+1,k_z}_{s,j,k_z}
-\frac{ieA}{m_0}\Big(\left(\mathcal{S}_{14}\right)^{s',j+1,k_z}_{s,j,k_z} - 
\left(\mathcal{S}_{23}\right)^{s',j+1,k_z}_{s,j,k_z}\Big)\Big],
\end{equation}
and 
\begin{equation}
(d_x-i d_y)^{s',j^\prime,k_z'}_{s,j,k_z}
=\delta_{k_z,k_z^\prime}\delta_{j',j-1}\Big[e 
\sum_{i=1}^{4}\left(\mathcal{R}_{ii}\right)^{s',j-1,k_z}_{s,j,k_z}
-\frac{ieA}{m_0}\Big(
\left(\mathcal{S}_{32}\right)^{s',j-1,k_z}_{s,j,k_z}
-\left(\mathcal{S}_{41}\right)^{s',j-1,k_z}_{s,j,k_z} \Big)\Big],
\end{equation}
where we have defined the overlap integrals
\begin{equation}
\left(\mathcal{S}_{mn}\right)_{s,j,k_z}^{s'j',k_{z}'}=\int_0^R d\rho \,\rho\, \Phi_{m,s',j',k'_z}^*(\rho)\Phi_{n,s,j,k_z}(\rho)
\end{equation}
and the matrix elements of radial position
\begin{equation}
\left(\mathcal{R}_{mn}\right)_{s,j,k_z}^{s'j',k_{z}'}=\int_0^R d\rho \,\rho^2\, \Phi_{m,s',j',k'_z}^*(\rho)\Phi_{n,s,j,k_z}(\rho).
\end{equation}
For circular polarization in the plane perpendicular to the nanowire axis, we find the conventional selection rule $j'=j\pm 1$, which is mandated by the
conservation of total-angular-momentum projection (including the photon's) parallel to the
nanowire axis. In addition, linear momentum $k_z$ parallel to the nanowire axis is conserved in any optical transition.
The energy threshold for absorption is associated with transitions between $(s'=+,\ j=\pm 1/2, 
k'_z=0)$ and $(s=-,\ j\mp 1/2, k_z=0)$. At the subband edge $(k_z=0$ and $k'_z=0)$ for $d_x+id_y$ only the overlap
integral $\left(\mathcal{S}_{14}\right)_{-,-1/2,0}^{+,1/2,0}$ is non-vanishing for absorption, 
while for emission the only non-vanishing overlap integral is $\left(\mathcal{S}_{23}\right)_{+,-1/2,0}^{-,1/2,0}$.  
For the opposite polarization, namely $d_x-id_y$, the non vanishing overlap integrals at the band edge are: $\left(\mathcal{S}_{32}\right)_{-,1/2,0}^{+,-1/2,0}$ for absorption and $\left(\mathcal{S}_{41}\right)_{+,1/2,0}^{-,-1/2,0}$ for emission, respectively. 
The overlap integrals relevant for the absorption threshold are shown in Fig.~\ref{fig:dipole1} asa function of the radius of the wire. 
It needs to be noticed that also the matrix elements of the radial position $\left(\mathcal{R}_{mn}\right)_{s,j,k_z}^{s'j',k_{z}'}$ contribute both to absorption and emission. The sum of these matrix elements for the case of  absorption is shown in Fig.~\ref{fig:dipole2} as a function of the radius of the wire. 
The finite radius of the nanowire does not affect the selection rules but leads to significative quantitative changes of the dipole matrix elements.

\begin{figure}
	\centering
	\includegraphics[width=\columnwidth]{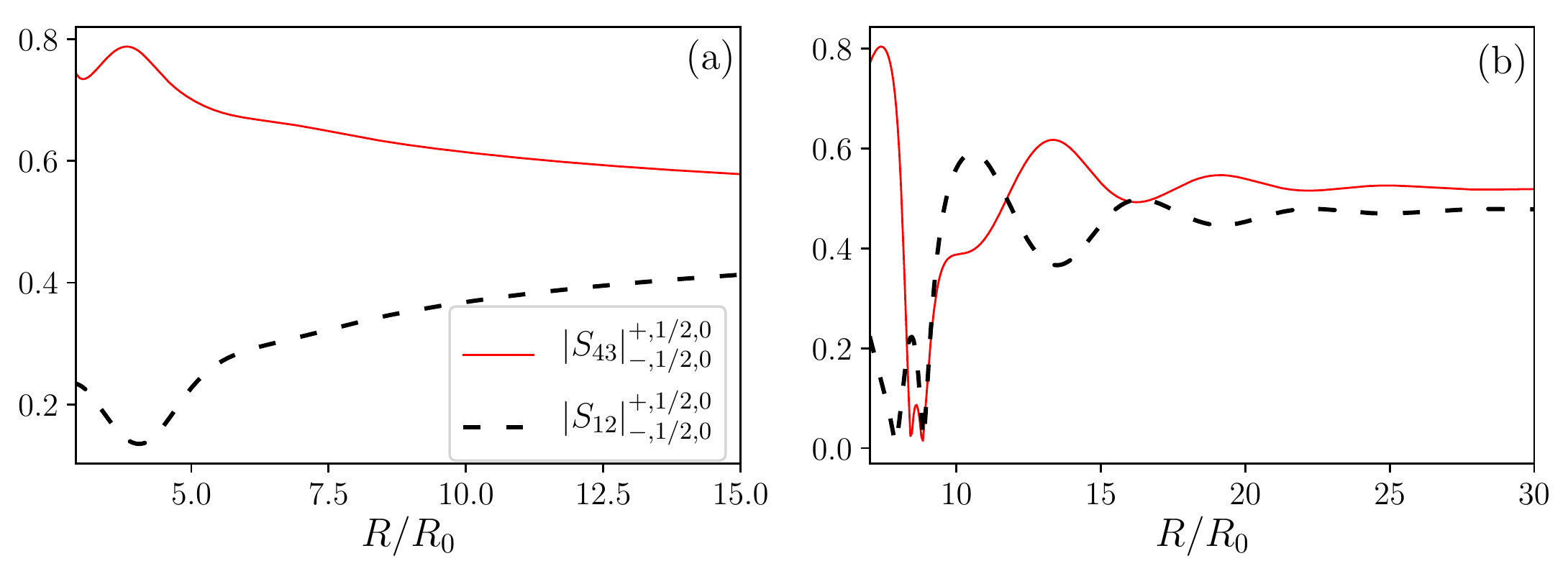}
\caption{Overlap integrals entering the dipole matrix element relevant for absorption for linearly-polarised (longitudinal) light as a function of the radius of the cylinder of (a) ${\rm Bi}_2{\rm Se}_3$ and (b) $\text{Bi}_2\text{Te}_3$ for $k_z=k_z'=0$. }
	\label{fig:dipole3}
\end{figure}

Matrix elements of the optical-dipole component parallel to the nanowire axis are given by
\begin{multline}\label{eq:paraDip}
(d_z)^{s'j',k_z'}_{s,j,k_z}=\Big\{e \sum_{i=1}^{4}\left(\mathcal{S}_{ii}\right)^{s',j,k'_z}_{s,j,k_z} \int dz \, z\, \frac{e^{i(k_z-k_z')z}}{2\pi}\\
+\delta_{k_z,k_z'}\frac{ieB}{2 m_0}
\left[ \left(\mathcal{S}_{21}\right)^{s',j,k_z}_{s,j,k_z} -
\left(\mathcal{S}_{12}\right)^{s',j,k_z}_{s,j,k_z} +
\left(\mathcal{S}_{34}\right)^{s',j,k_z}_{s,j,k_z} -
\left(\mathcal{S}_{43}\right)^{s',j,k_z}_{s,j,k_z} \right]
\Big\}\delta_{j',j}.
\end{multline}
The first term on the r.h.s.\ of Eq.~(\ref{eq:paraDip}) is ill-defined because the
envelope functions are not localized in their dependence on the $z$ coordinate
and, hence, the dipole approximation is not valid. However, the remaining
basis-function-mediated contributions describe valid optical transitions. For these, both linear momentum $k_z$ and the total-angular-momentum
projection $j$
parallel to the nanowire axis are the same for initial and final states involved in
optical transitions. For states at the energy threshold of absorption, we find that the only non vanishing overlap integrals are $\left(\mathcal{S}_{12}\right)_{-,\pm 1/2,0}^{+,\pm 1/2,0}$ and $\left(\mathcal{S}_{43}\right)_{-,\pm 1/2,0}^{+,\pm 1/2,0}$, while for emission the non vanishing overlap integrals are $\left(\mathcal{S}_{21}\right)_{+,\pm 1/2,0}^{-,\pm 1/2,0}$ and $\left(\mathcal{S}_{34}\right)_{+,\pm 1/2,0}^{-,\pm 1/2,0}$. The overlap integrals relevant for absorption are shown in Fig.~\ref{fig:dipole3}. 
Again, the selection rules for optical transitions are consistent with the basic
symmetries associated with a cylindrical-nanowire geometry, and finite-size
effects are manifested as significant quantitative changes in the magnitude of
dipole matrix elements.
\section{Conclusions}
In this paper we have studied a nanowire made of TI. In particular, we have
provided the analytical form of the energy eigenfuctions, which is central to the derivation of an analytical secular equation for the eigenenergies. This secular equation, on one hand, enables an analytical expansion for large radii and, on the other hand, is amenable to straightforward numerical solution. We study the dependence of the eigenenergies on the radius of the wire and we find oscillations as a function of the radius, which are very pronounced for Bi$_2$Te$_3$. 
The analytical form of the energy eigenfuctions enables the computation of the matrix elements of any single-particle operator. We have considered the optical dipole matrix elements. While we find the usual selection rules for absorption/emission, the value of the matrix elements is strongly dependent on the radius of the cylinder. 
\appendix
\section{Secular equation for confined states}
\label{secul}
In this Appendix we provide the detailed derivation of the secular equation for the state of the TI cylinder. 
Acting with the Hamiltonian~(\ref{eq:Hsum}) on the wave function Eq.~(\ref{eq:Ansatz1}) and looking for eigenfunctions with energy $E$, we obtain
\begin{equation}
\label{eq:eigen1}
\left(
\begin{smallmatrix}
m_\perp+ m_{-}(j,k_z)-E & B k_z & 0 & -i A \left[\pr +\frac{1}{\rho} \left(j+\frac{1}{2}\right)\right] \\
B k_z& - \left[ m_\perp+ m_{-}(j,k_z)+E\right] &  -i A \left[\pr +\frac{1}{\rho} \left(j+\frac{1}{2}\right)\right] & 0\\
0 &  -i A \left[\pr -\frac{1}{\rho} \left(j-\frac{1}{2}\right)\right] & m_\perp+ m_{+}(j,k_z)-E & -B k_z\\
-i A \left[\pr -\frac{1}{\rho} \left(j-\frac{1}{2}\right)\right]   & 0 & -B k_z & -\left[m_\perp+ m_{+}(j,k_z) +E\right]
\end{smallmatrix}
\right)
\left(\begin{matrix}
\Phi_1(\rho)\\
\Phi_2(\rho)\\
\Phi_3(\rho)\\
\Phi_4(\rho)
\end{matrix}
\right)
=0,
\end{equation}
where we have defined $m_{\pm}(j,k_z)=m_2\frac{1}{\rho^2}\left(j\pm\frac{1}{2}\right)^2+m_1 k_z^2$.
To solve the eigensystem Eq.~(\ref{eq:eigen1}) we make the \textit{Ans\"atz}
\begin{align}
\label{eq:Ansatz-app}
\left(\begin{matrix}
\Phi_1(\rho)\\
\Phi_2(\rho)\\
\Phi_3(\rho)\\
\Phi_4(\rho)
\end{matrix}
\right)=
\left(
\begin{matrix}
c_1 J_{j-\frac{1}{2}}(\kappa\rho)\\
c_2 J_{j-\frac{1}{2}}(\kappa\rho)\\
c_3 J_{j+\frac{1}{2}}(\kappa\rho)\\
c_4 J_{j+\frac{1}{2}}(\kappa\rho)
\end{matrix}
\right),
\end{align}
where $J_n(z)$ is a Bessel function of the first kind and $\kappa$ and the coefficients $c_1,\dots, c_4$ need to be determined. Substituting the \textit{Ans\"atz} Eq.~(\ref{eq:Ansatz-app}) in (\ref{eq:eigen1}), we obtain the following equation for the coefficients
\begin{align}
\label{eq:eigen2}
\left(
\begin{smallmatrix}
-\left(\kappa^2 +\frac{m_1}{m_2} k_z^2+\frac{m_0-E}{m_2}\right) & -\frac{B k_z}{m_2} & 0 & i\frac{A \kappa}{m_2} \\
\frac{B k_z}{m_2}& -\left(\kappa^2 +\frac{m_1}{m_2}  k_z^2+\frac{m_0+E}{m_2}\right) & -i\frac{A \kappa}{m_2}  & 0\\
0 & -i\frac{A \kappa}{m_2}  & -\left(\kappa^2 +\frac{m_1}{m_2} k_z^2+\frac{m_0-E}{m_2}\right) & \frac{B k_z}{m_2} \\
i\frac{A \kappa}{m_2}  & 0 &- \frac{B k_z}{m_2} &-\left(\kappa^2 +\frac{m_1}{m_2} k_z^2+\frac{m_0+E}{m_2}\right) \\
\end{smallmatrix}
\right)
\left(\begin{matrix}
c_1\\
c_2\\
c_3\\
c_4
\end{matrix}
\right)
=0.
\end{align}

Equation~(\ref{eq:eigen2}) has non trivial solutions for 
\begin{align}
\left(\kappa^2+\frac{m_1}{m_2} k_z^2+\frac{m_0}{m_2} \right)^2+\frac{A^2}{m_2^2}\kappa^2+\frac{B^2}{m_2^2} k_z^2-\frac{E^2}{m_2^2}=0
\end{align}
which yields\footnote{The negative sign for the outer square root does not give a different solution and therefore should not be considered due to the property of the Bessel's functions: $J_n(z)=(-1)^n J_n(-z)$ for integer $n$.}
\begin{align}
\kappa=\kappa_\pm=\sqrt{-\left(\frac{m_0}{m_2}+\frac{A^2}{2 m_2^2}+\frac{m_1}{m_2} k_z^2\right)\pm
\sqrt{\frac{A^4}{4 m_2^4}+\frac{E^2}{m_2^2}+\frac{A^2 m_0}{m_2^3}+\left(\frac{A^2}{m_2^2}\frac{m_1}{m_2} -\frac{B^2}{m_2^2}\right)k_z^2}.
}
\end{align}
There are four independent solutions for $(c_1,c_2,c_3,c_4)^{T}$ and are given by
\begin{align}
& \left(
\frac{i A \kappa_\pm}{\Delta_\pm},\
0,\ \frac{ B k_z}{\Delta_\pm},\ 1
\right)^{T},\\
&\left(
-\frac{B k_z}{\Delta_\pm},\ 
1,\ -\frac{i A \kappa_\pm}{\Delta_\pm},\ 0
\right)^{T},
\end{align}
where we have introduced the following abbreviation $\Delta_\pm=m_2\kappa_\pm^2+m_1 k_z^2+m_0-E$.
The general solution with quantum numbers $k_z$, $j$ and $E$ can therefore be written as
\begin{align}
\Psi(\rho,\varphi,z)=\frac{e^{i k_z z}}{\sqrt{2\pi}}\sum_{s=\pm}\left\{
\alpha_s\left(
\begin{smallmatrix}
\frac{i A \kappa_s}{\Delta_s}\ J_{j-\frac{1}{2}}(\kappa_s\rho)\ e^{i(j-\frac{1}{2})\varphi}\\
0\\ \frac{ B k_z}{\Delta_s}\ J_{j+\frac{1}{2}}(\kappa_s\rho) \ e^{i(j+\frac{1}{2})\varphi}\\ J_{j+\frac{1}{2}}(\kappa_s\rho) \ e^{i(j+\frac{1}{2})\varphi}
\end{smallmatrix}\right)+\beta_s \left(
\begin{smallmatrix}
- \frac{ B k_z}{\Delta_s}\ J_{j-\frac{1}{2}}(\kappa_s\rho)\ e^{i(j-\frac{1}{2})\varphi}\\
J_{j-\frac{1}{2}}(\kappa_s\rho)\ e^{i(j-\frac{1}{2})\varphi}\\ -\frac{ i A \kappa_s}{\Delta_s}\ J_{j+\frac{1}{2}}(\kappa_s\rho) \ e^{i(j+\frac{1}{2})\varphi}\\0
\end{smallmatrix}\right)
\right\}.
\end{align}
 Assuming a hard-wall cylindrical confinement potential of radius $R$, we need to impose the boundary condition $\Psi(R,\varphi,z)=0$ which leads to the following system of equations:
\begin{multline}
\left(
\begin{matrix}
\frac{i A \kappa_+}{\Delta_+}\ J_{j-\frac{1}{2}}(\kappa_+ R) &
- \frac{ B k_z}{\Delta_+}\ J_{j-\frac{1}{2}}(\kappa_+ R) & \frac{i A \kappa_-}{\Delta_-}\ J_{j-\frac{1}{2}}(\kappa_- R) &
- \frac{ B k_z}{\Delta_-}\ J_{j-\frac{1}{2}}(\kappa_- R)
\\
0 & J_{j-\frac{1}{2}}(\kappa_+ R) & 0 & J_{j-\frac{1}{2}}(\kappa_- R) 
\\ 
\frac{ B k_z}{\Delta_+}\ J_{j+\frac{1}{2}}(\kappa_+ R) & 
-\frac{ i A \kappa_+}{\Delta_+}\ J_{j+\frac{1}{2}}(\kappa_+R) &
\frac{ B k_z}{\Delta_-}\ J_{j+\frac{1}{2}}(\kappa_- R) & 
-\frac{ i A \kappa_-}{\Delta_-}\ J_{j+\frac{1}{2}}(\kappa_- R) 
\\ J_{j+\frac{1}{2}}(\kappa_+ R) & 0& J_{j+\frac{1}{2}}(\kappa_-R) & 0
\end{matrix}\right)
\label{eq:eigen3}
\left(\begin{matrix}
\alpha_+\\
\beta_+\\
\alpha_-\\
\beta_-
\end{matrix}
\right)\\
=0.
\end{multline}
 
We then obtain the secular equation
\begin{align}
\nonumber
& \left[\kappa_+ \Delta_- J_{j-\frac{1}{2}}(\kappa_- R)J_{j+\frac{1}{2}}(\kappa_+ R)
-\kappa_- \Delta_+ J_{j-\frac{1}{2}}(\kappa_+ R)J_{j+\frac{1}{2}}(\kappa_- R)
\right] \times\\
\nonumber & 
\left[\kappa_+ \Delta_- J_{j-\frac{1}{2}}(\kappa_+ R)J_{j+\frac{1}{2}}(\kappa_- R)
-\kappa_- \Delta_+ J_{j-\frac{1}{2}}(\kappa_- R)J_{j+\frac{1}{2}}(\kappa_+ R)
\right] +\\
\label{eq:secularfinal}
&\frac{B^2}{A^2} k_z^2(\Delta_{+}-\Delta_{-})^2 J_{j-\frac{1}{2}}(\kappa_- R)J_{j-\frac{1}{2}}(\kappa_+ R)J_{j+\frac{1}{2}}(\kappa_- R)J_{j+\frac{1}{2}}(\kappa_+ R)
=0.
\end{align}
Notice that the term in the third line of Eq.~(\ref{eq:secularfinal}) vanishes for $k_z=0$.
By simple algebraic manipulations, Eq.~(\ref{eq:secularfinal}) can be cast in the form of Eq.~(\ref{seceq2}). 

\section{Small-radius limit}
\label{small}
\begin{figure}
	\centering
	\includegraphics[width=0.6\columnwidth]{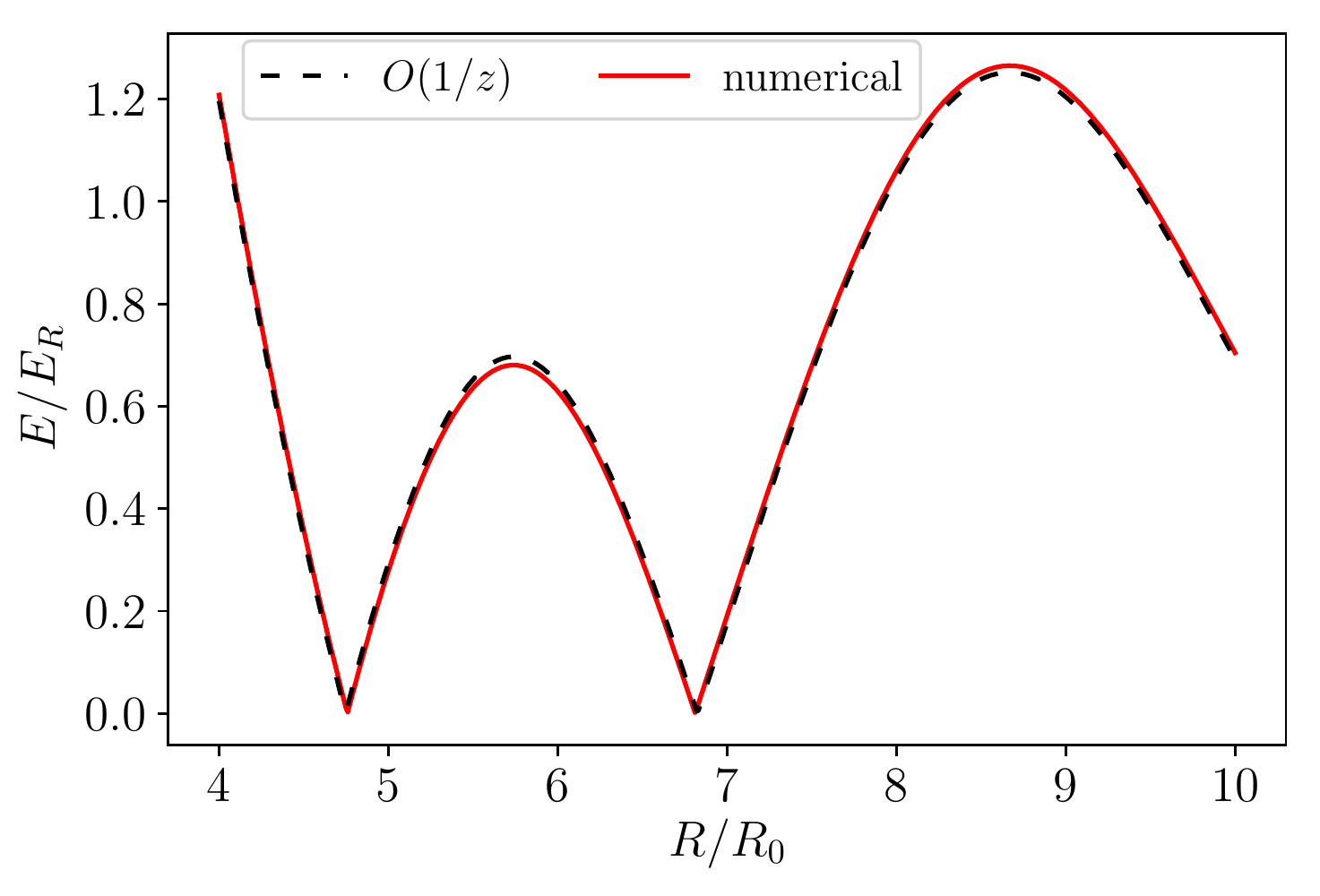}
\caption{Eigenenergies with $j=\frac{1}{2}$ and $s=+$ for a cylinder of $\text{Bi}_2\text{Te}_3$ as a function of radius for $k_z=0$. The solid red curve is the exact numerical solution of Eq.~(\ref{seceq2}) while the dashed black curves is obtained by means of the Hankel's expansion at first order in $1/z$. }
	\label{fig:evsrsmallrad}
\end{figure}
In section~\ref{numerical} we found interesting finite-size effects for small values of the radius $R$, such as the oscillatory behaviour of the eigenenergies.
In order to understand the origin of  the oscillations in Fig.~\ref{fig:evsr}, here we use the Hankel's asymptotic expansion [Eq.~(\ref{eq:hankel})], but without approximating the trigonometric functions, and solve the secular equation at each given order. 
The plot of the eigenenergy as a function of $R$, obtained by taking into account only the first order in $1/z$ [$P(n,z)=1$ and $Q(n,z)=(4 n^2-1)/(8z)$], is shown in Fig.~\ref{fig:evsrsmallrad} as a dashed black curve: it is found to agree remarkably well with the full numerical results (solid red curve). The expansion up to second order in $1/z$ (not shown) [$P(n,z)=1-{(4n^2-1)(4n^2-9)}/{(128z^2)}$ and $Q(n,z)=(4 n^2-1)/(8z)$] is practically indistinguishable from the full numerical results.

\end{document}